\documentclass[12pt,preprint]{aastex}

\begin{document}
\title{First R I lights and their photometric analyses of GSC 02393-00680 }

\author{Liu L.\altaffilmark{1,2,3}, Qian S.-B.\altaffilmark{1,2,3}, He J.-J.\altaffilmark{1,2,3},
Li L.-J.\altaffilmark{1,2,3} and Liao W.-P.\altaffilmark{1,2,3}}

\altaffiltext{1}{National Astronomical Observatories/Yunnan
Observatory, Chinese Academy of Sciences, \\P.O. Box 110, 650011
Kunming, P.R. China\\ (e-mail: creator\_ll.student@sina.com;
LiuL@ynao.ac.cn)}

\altaffiltext{2}{United Laboratory of Optical Astronomy, Chinese
Academy of Sciences (ULOAC),\\ 100012 Beijing, P. R. China}

\altaffiltext{3}{Graduate School of the Chinese Academy of
Sciences,\\ 100012 Beijing, P. R. China }


\begin{abstract}

We obtained complete $R$ and $I$ light curves of GSC 02393-00680 in
2008 and analyzed them with the 2003 version of the W-D code. It is
shown that GSC 02393-00680 is a W-type shallow contact binary system
with a high mass ratio $q=1.600$ and a degree of contact factor
$f=5.0\%(\pm1.3\%)$. It will be a good example to check up on the
TRO theory. A period investigation based on all available data
suggests that the system has a small-amplitude period oscillation
($A_3=0.^{d}0030$; $T_3=1.92$years). This may indicate it has a
moderate mass close third body, which is similar to XY Leo.

\end{abstract}

\section{Introduction}

GSC 02393-00680 is a newly discovered EW-type binary in recent
years. Its times of minima were reported by Martignoni (2006) and
Bl\"attler \& Diethelm (2006), respectively. Martignoni (2006) have
got 10 times of minima and a linear epoch $2453354.396+0.31657\times
E$. Meanwhile, having used his 15cm Starfire refractor without
filter to observe this object on 7 nights, Bl\"attler and Diethelm
(2006) obtained 15 times of minima and yielded a linear epoch
$2453686.4414+0.316535\times E$. They also published a light curve
of the binary, in which we can get some information such as
magnitude amplitude (about 0.3mag), the difference in the depths of
the two eclipses. However, the data in their light curve are
scattered to some extent, and it is not clear that the light curve
shows the O'Connell effect (O'Connell, 1951). We found the effect in
our observed light curves and adopt a dark spot model to interpret
it. The aim of the present study is to analyse our $R$ and $I$ light
curves to get much information of this new eclipsing binary.

\section{Observations in $R$ and $I$ Band}

GSC 02393-00680 was firstly observed in December 27, 2007 with the
PI512 TKB CCD photometric system attached to the 60cm reflecting
telescope at the Yunnan Observatory in China. However, the quality
of the data was not approving, so we observed it again in January 6,
2008 with the PI1024 TKB CCD photometric system attached to the 1m
reflecting telescope at the Yunnan Observatory in China. The $R$ and
$I$ color systems used in the latter instrument are close to the
standard $UBVRI$ system. The effective field of view of the
photometric system is $6.5\times6.5$ arcmin at the Cassegrain focus.
The integration time for each image is 20\,s. The comparison star
chosen was 2MASS 05085699+3204216 and the check star was 2MASS
05083765+3205108. These stars are as bright as the target and close
to the target (less than 4 arcmin) so that the extinction conditions
of them were the same. PHOT (measure magnitudes for a list of stars)
of the aperture photometry package of IRAF was used to reduce the
observed images. Through the observation we obtained complete $R$
and $I$ lights. By calculating the phase of the observations with
the equation $2454472.0846+0.316535\times E$, the light curves are
plotted (figure 1). The original data in the $R$ and $I$ bands are
listed in table 1 and 2. It is shown that the data are of high
quality and the light variation is typical of EW. The lights of the
comparison subtract the check star are straight lines which strongly
proved the chosen comparison star was a light constancy one and the
light changes truly came from the target. Therefore, extinction
correction was not made. A parabolic fitting was used to determine
the times of light minima with the least square-method, since the
light minimum is symmetric. In all, four new epochs of light minima
were obtained in our observations. They were listed in the last of
table 3.

\section{Orbital period variations}
Though the period of the investigation of this binary is not long,
the obtained data already have reflected some rules of its period
variations. Having collected all useful times of minima, we correct
the linear ephemeris as:

\begin{eqnarray}
{\rm Min.~I}&=&2453686.4425(\pm0.0005)\nonumber\\
&&+0^{d}.31653591(\pm0.00000040)\times{E}
\end{eqnarray}

\noindent The $(O-C)_1$ values with respect to the linear ephemeris
are listed in the sixth column of table 3. The corresponding
$(O-C)_1$ diagram is displayed in figure 2. It is no doubt that the
general trend of GSC 02393-00680, shown in figure 3, is a cyclic
variation.

The general $(O-C)_{1}$ trend of GSC 02393-00680, shown in figure 2,
indicates the cyclic period changes. Beside this, there is possibly
very clear evidence that it has a small-amplitude period
oscillation. Assuming that the period oscillation is cyclic, then,
based on a least-square method, a sinusoidal term is added to a
linear ephemeris to give a better fit to the $(O-C)_{1}$ curve
(solid line in figure 2). After having adjusted the period of the
cyclic change for several times, we found the best value is $\omega
=0^{\circ}.1625$. We fixed this value and simulated the rest
coefficients. The final result is
\begin{eqnarray}
{\rm Min.~I} &=&2453686.4428(\pm0.0004)\nonumber\\
&&+0^{d}.31653708(\pm0.00000061)\times{E}\nonumber\\
&&+0.0030(\pm0.0013)\sin(0^{\circ}.1625E+199.6(\pm0.3))
\end{eqnarray}
 \noindent The $(O-C)_{2}$ values with respect to the cyclic ephemeris in
equation (2) are shown in figure 3, and the corresponding residuals
were plotted in the lower panel of figure 2. From it we can see
residuals are in a horizontal line on average, except some scatters.
By using this relation,
\begin{equation}
\omega=360^{\circ}P_{e}/T
\end{equation}

\noindent the period of the orbital period oscillation is determined
to be T=1.92 years. This may suggest that there exists a
small-amplitude oscillation ($A_3=0.^{d}0030$) in the period changes
which can be explained by the presence of an third body in the
system.

\section{Photometric solutions}
We give the photometric parameters of GSC 02393-00680 first. To get
an initial value of mass ratio q, a q-search method with the 2003
version of the W-D program (Wilson \& Devinney, 1971, Wilson, 1990,
1994, Wilson \& Van Hamme, 2003) was used (figure 4). We fixed q to
0.3, 0.4, 0.5 and so on, as figure 4 shows. It can be seen that the
best value is around $q=1.61$.

During the solution, the temperature of star 1 (star eclipsed at
primary light minimum) was fixed at $T_1=5860$K. This is estimated
from its $H-J$ color index (Cutri et al. 2003; Cox 2000). The
bolometric albedo $A_1=A_2=0.5$ (Rucinski 1969) and the values of
the gravity-darkening coefficient $g_1=g_2=0.32$ (Lucy 1967) were
used, which correspond to the common convective envelope of both
components. According to Claret \& Gimenez (1990), the root
limb-darkening coefficients were used (table 4). We adjusted the
orbital inclination $i$; the mean temperature of star 2, $T_2$; the
monochromatic luminosity of star 1, $L_{1R}$ and the dimensionless
potential of star 1 ($\Omega_1=\Omega_2$, mode 3 for contact
configuration). A small O'Connell effect of the system can not be
ignored. GSC 02393-00680 might be a sun-like star. So it seems
probable that starspots appeared on the surface of the star. In
fact, as many researchers have done before (e.g., Binnedijk, 1960,
Mullan, 1975, Bell, et al., 1990, Linnell \& Olson, 1989), we add a
spot on the star 2, which proved in the following as the cooler more
massive primary component. The photometric solutions are listed in
table 4 and the theoretical light curves computed with those
photometric elements are plotted in figure 5. In table 4, $\theta$
is the latitude of the star spot center, measured like our earth
from $+90$ degrees at the north pole to $-90$ degrees at the south
pole. $\psi$ is the longitude of a star spot center, measured
counter-clockwise (as viewed from above the +z axis) from the line
of star centers from 0 to 360 degrees. $\Omega$ is the angular
radius of a star spot, in degrees. $T_s/T_*$ is the temperature
factor of a spot, that specifies the ratio of local spot temperature
to local temperature which would obtain without the spot. In order
to get an image of the binary and the spot on it in our mind, the
geometrical structure of GSC 02393-00680 is displayed in figure 6.

The difference of temperatures between the two components according
to our solution is 500 K. That may indicate that the system did not
reach thermal contact (Lucy \& Wilson 1979, Rucinski \& Duerbeck
1997). In that case, the O'Connell effect could be caused by a hot
spot yielded by a mass transfer from the less massive component to
the more one. However, the outcome is that the simulated hot spot
presents at a fallacious position which is inadequate for the
physical laws. If it be forced a hot spot, it can only present on
the small component. This is quite improbable. So we only give the
cool spot solution. The suspected third light which was able to
affect the whole light curves was also considered in our simulations
We adjusted the parameter $l_3$ in the W-D code, but found that the
numerical third light calculated by the program is negative. That
tells us the luminosity contribution of third light is not obvious
in the system. Hence, the final results did not include the third
light.

\section{Discussions and conclusions}

The orbital period was revised as 0.31653619 days by using the 31
CCD timings of GSC 02393-00680 listed in table 3. The system is a
high mass ratio shallow contact W-type binary with $q=1.600$,
$f=5.0\%$. These suggest that the system may be a marginal contact
binary. Marginal contact binaries, whose filling factors are very
small ($f\leqslant 10\%$), are indicators of evolution time scale
into the contact stage. So far there are several examples of
discovered marginal contacts, such as II CMa (Liu et al. 2008), V803
Aql (Samec et al. 1993), FG Sct (Bradstreet 1985), RW PsA (Lucy \&
Wilson 1979), XZ Leo (Niarchos et al. 1994), S Ant (Russo et al.
1982), TW Cet (Russo et al. 1982) , AB And, GZ And (Baran et al.
2004) , HT Vir (Zola et al. 2005) , XY Leo (Djura\v{s}evi\'{c} et
al. 2006, Rucinski et al. 2007), V508 Oph (Lapasset \& Gomez 1990).
All of these binaries, at least, has a common nature; namely their
contact degrees are less than $10\%$. Markedly, this system is quite
similar to the well studied system XY Leo. We will discuss them
below.

High mass ratio, low contact factor contact systems are the key to
understanding the evolution status of a close binary from the near
contact phase to the contact phase. Almost all the acceptable models
of contact binary predict that as the age in the contact phase is
increasing, the mass ratio $q$ is decreasing. This is due to
ineluctable mass transfer from the less massive component to the
more massive one when the system came into the contact stage. The
reasons are the system is a W-type system; the present less massive
component evolved faster than the present more massive one; the mass
transfer began from the less massive component. The mass ratio and
the low contact factor suggest that GSC 02393-00680 has just come
into contact and that it is young as a contact system. On the other
hand, it might also be near to the broken-contact phase. This
depends on the direction of the period change, i.e., increase or
decrease. If the period increases, the system is going to the
broken-contact phase; if the period decreases, it has just come into
the contact configuration. Which one is the truth? The both case may
be possible. A further study is needed to answer this question.

Several authors (e.g. Lucy 1976, Flannery 1976, Robertson \&
Eggleton 1977) have proposed thermal non-equilibrium models which
assumed that the components of a contact binary are not in thermal
equilibrium and can exchange matter freely between them (the thermal
relaxation oscillation model: TRO model). They predicted that the
systems must undergo cycles around the state of marginal contact if
the total mass and angular momentum are conserved. TRO model can
explain the shallow contact feature observed for many W UMa-type
stars and predict that many parameters, such as the separation, the
mass ratio and the orbital period, should vary on thermal
time-scales and, indeed, secular period changes (both increasing and
decreasing) having time-scales of the right order of magnitude are
observed. But there are still some difficulties of TRO model. To
avoid them, Rahunen (1981) dropped the usual assumption of
conservation of angular momentum. He treated the orbital angular
momentum as a free parameter, keeping the system in marginal contact
at each time-step. He concluded that when the angular momentum loss
(AML) rate is $d\textrm{ln} j/dt\sim2{\times}10^{-9} yr^{-1}$,
contact binaries can keep in shallow contact phase. A large AML rate
can cause the system to quickly reach the L2-surface and it may
finally coalesce into a single star. On the contrary, a smaller AML
rate than the critical rate cannot maintain a system in contact and
the system oscillates again. But this model leaves two questions
unsolved, as pointed by Rahunen (1981) and by Smith (1984): what is
the mechanism of AML and what mechanism keeps the system permanently
in shallow contact? Qian (2001a, b, 2003a) have shown an
evolutionary scenario of contact binary stars. According to this
scenario, the evolution of a contact binary may be the combination
of the thermal relaxation oscillation (TRO) and the variable angular
momentum loss (AML) via the change of depth of contact. Systems
(e.g., V417 Aql, see Qian 2003b) with a secular decreasing period
are on the AML-controlled stage, while those (e.g., CE Leo, see Qian
2002) showing an increasing period are on the TRO-controlled stage.
GSC 02393-00680 will to be a good target to check up on the TRO
theory.

The alternate period change of a close binary containing at least
one solar-type component can be interpreted by the mechanism of
magnetic activity (e.g., Applegate 1992, Lanza et al. 1998).
However, for contact binary stars, we do not sure whether this
mechanism will work. Moreover we do not know how it might work. We
think the period oscillation may be caused by the light-time effect
of a tertiary component. As we can see from figure 2, 3, the trend
of O-C variation is clear. In section 3, a theoretical solution of
the orbit for the assumed tertiary star was calculated. By using
this equation:
\begin{equation}
f(m)=\frac{4\pi^{2}} {{\it
G}T_3^{2}}\times(a_{12}^{\prime}\sin{i}^{\prime})^{3},
\end{equation}
where $a_{12}^{\prime}\sin{i}^{\prime}=A_3\times{c}$ (where c is the
speed of light), the mass function from the tertiary component is
computed. Then, with the following equation:
\begin{equation}
f(m)=\frac{(M_{3}\sin{i^{\prime}})^{3}} {(M_{1}+M_{2}+M_{3})^{2}}.
\end{equation}
For estimation, taking the assumed physical parameters
$M_1=0.6M_{\odot}$, $M_2=1.0M_{\odot}$ (A typical sun-like star),
the masses and the orbital radii of the third companion are
computed. The values of the masses and the orbital radii of the
third component stars for several different orbital inclinations
($i^{\prime}$) are shown in table 5. If the tertiary companion is
coplanar to the eclipsing pair (i.e.,with the same inclination as
the eclipsing binary), its mass should be $m_3=0.67M_{\odot}$.

Based on all available eclipse times, the period changes of the
contact binary star were discussed in the previous section. Because
the period of the investigation of this binary is not long, the
long-term period changes which expected to be revealed by general
O-C trend are not clear. It needs further observations. Meanwhile, a
small-amplitude period oscillation ($A_3=0.^{d}0030$) was suggested
in the period changes. This cyclic variation is due to either the
light-time effect via the presence of an assumed third body or the
variation of the gravitational quadruple momentum via magnetic
activity cycle of the cool component. Whatever the explanation is,
the system is very interesting because of the possible
small-amplitude and fast ($P_3$=1.92yr) period oscillation. The
system is very similar to XY Leo, see table 6. In the case of XY
Leo, the third body in this system is assured. It confirm that a
short period moderate mass third body indeed exists. It is possible
to be a cool star which do not contribute luminosity to the system,
otherwise we can get some information about the third light from the
W-D code. Of course we expect a spectral analysis. In every point it
is worth to investigate the system further.

\vskip 0.3in \noindent At first, We would like to give our great
thanks to the anonymous referee who had given us a lot of
constructive suggestions which had enormously improved the paper.
This work is partly supported by Yunnan Natural Science Foundation
Foundation (No. 2005A0059M), Chinese Natural Science Foundation
(No.10573032, No.10433030, and No.10573013), The Ministry of Science
and Technology of the People¡¯s Republic of China through grant
2007CB815406 and The Chinese Academy of Sciences grant No.
O8ZKY11001. New observations of the system were obtained with the
60cm and 1-m telescope at Yunnan Observatory.

\begin{table}
\begin{footnotesize}
\caption{CCD photometric data of GSC 02393-00680 in I band obtained
on January 6, 2008.}
\begin{tabular}{lllllllll}\hline\hline
JD (Hel.)    & Phase & $\Delta{m}$ & JD (Hel.)    & Phase &
$\Delta{m}$ & JD (Hel.)    & Phase & $\Delta{m}$ \\
2454400+   &             &            &2454400+     &            &
&2454400+ &              &            \\\hline
71.9744   & 0.6518  &  0.725  & 72.0725   & 0.9617 &  0.897  &  72.1793  &  0.2991  &  0.659  \\
71.9763   & 0.6578  &  0.707  & 72.0749   & 0.9693 &  0.902  &  72.1812  &  0.3051  &  0.664  \\
71.9782   & 0.6638  &  0.714  & 72.0771   & 0.9762 &  0.915  &  72.1831  &  0.3111  &  0.667  \\
71.9799   & 0.6691  &  0.710  & 72.0793   & 0.9832 &  0.922  &  72.1853  &  0.3180  &  0.665  \\
71.9823   & 0.6767  &  0.703  & 72.0819   & 0.9914 &  0.929  &  72.1877  &  0.3256  &  0.675  \\
71.9842   & 0.6827  &  0.688  & 72.0846   & 0.9999 &  0.921  &  72.1898  &  0.3323  &  0.679  \\
71.9860   & 0.6884  &  0.689  & 72.0866   & 0.0062 &  0.926  &  72.1941  &  0.3458  &  0.697  \\
71.9879   & 0.6944  &  0.676  & 72.0886   & 0.0125 &  0.918  &  72.1965  &  0.3534  &  0.713  \\
71.9899   & 0.7007  &  0.676  & 72.0909   & 0.0198 &  0.918  &  72.1985  &  0.3597  &  0.707  \\
71.9918   & 0.7067  &  0.685  & 72.0930   & 0.0264 &  0.913  &  72.2010  &  0.3676  &  0.720  \\
71.9936   & 0.7124  &  0.676  & 72.0950   & 0.0328 &  0.902  &  72.2030  &  0.3740  &  0.731  \\
71.9955   & 0.7184  &  0.674  & 72.0969   & 0.0388 &  0.892  &  72.2050  &  0.3803  &  0.741  \\
71.9973   & 0.7241  &  0.672  & 72.0990   & 0.0454 &  0.877  &  72.2070  &  0.3866  &  0.749  \\
72.0006   & 0.7345  &  0.666  & 72.1010   & 0.0517 &  0.868  &  72.2097  &  0.3951  &  0.759  \\
72.0027   & 0.7412  &  0.667  & 72.1030   & 0.0580 &  0.860  &  72.2123  &  0.4033  &  0.762  \\
72.0047   & 0.7475  &  0.666  & 72.1050   & 0.0644 &  0.841  &  72.2148  &  0.4112  &  0.779  \\
72.0066   & 0.7535  &  0.672  & 72.1075   & 0.0723 &  0.830  &  72.2167  &  0.4172  &  0.783  \\
72.0085   & 0.7595  &  0.663  & 72.1096   & 0.0789 &  0.818  &  72.2201  &  0.4280  &  0.793  \\
72.0103   & 0.7652  &  0.671  & 72.1122   & 0.0871 &  0.803  &  72.2223  &  0.4349  &  0.804  \\
72.0121   & 0.7709  &  0.679  & 72.1141   & 0.0931 &  0.787  &  72.2257  &  0.4457  &  0.832  \\
72.0141   & 0.7772  &  0.669  & 72.1162   & 0.0997 &  0.774  &  72.2289  &  0.4558  &  0.847  \\
72.0162   & 0.7838  &  0.681  & 72.1194   & 0.1098 &  0.760  &  72.2311  &  0.4627  &  0.856  \\
72.0181   & 0.7898  &  0.681  & 72.1220   & 0.1181 &  0.754  &  72.2339  &  0.4716  &  0.874  \\
72.0201   & 0.7961  &  0.684  & 72.1258   & 0.1301 &  0.742  &  72.2362  &  0.4788  &  0.883  \\
72.0221   & 0.8025  &  0.690  & 72.1267   & 0.1329 &  0.736  &  72.2388  &  0.4871  &  0.882  \\
72.0243   & 0.8094  &  0.697  & 72.1274   & 0.1351 &  0.734  &  72.2413  &  0.4950  &  0.882  \\
72.0265   & 0.8164  &  0.701  & 72.1292   & 0.1408 &  0.723  &  72.2444  &  0.5047  &  0.880  \\
72.0285   & 0.8227  &  0.708  & 72.1312   & 0.1471 &  0.722  &  72.2466  &  0.5117  &  0.884  \\
72.0321   & 0.8341  &  0.716  & 72.1331   & 0.1531 &  0.709  &  72.2488  &  0.5187  &  0.879  \\
72.0340   & 0.8401  &  0.720  & 72.1350   & 0.1591 &  0.707  &  72.2510  &  0.5256  &  0.885  \\
72.0362   & 0.8470  &  0.726  & 72.1404   & 0.1762 &  0.692  &  72.2534  &  0.5332  &  0.876  \\
72.0383   & 0.8536  &  0.733  & 72.1425   & 0.1828 &  0.689  &  72.2559  &  0.5411  &  0.869  \\
72.0406   & 0.8609  &  0.741  & 72.1445   & 0.1891 &  0.686  &  72.2583  &  0.5487  &  0.857  \\
72.0426   & 0.8672  &  0.747  & 72.1467   & 0.1961 &  0.674  &  72.2607  &  0.5562  &  0.841  \\
72.0446   & 0.8735  &  0.749  & 72.1491   & 0.2037 &  0.673  &  72.2633  &  0.5645  &  0.837  \\
72.0466   & 0.8799  &  0.755  & 72.1509   & 0.2094 &  0.665  &  72.2658  &  0.5724  &  0.830  \\
72.0487   & 0.8865  &  0.767  & 72.1551   & 0.2226 &  0.654  &  72.2690  &  0.5825  &  0.797  \\
72.0507   & 0.8928  &  0.773  & 72.1570   & 0.2286 &  0.657  &  72.2714  &  0.5900  &  0.794  \\
72.0526   & 0.8988  &  0.782  & 72.1591   & 0.2353 &  0.652  &  72.2735  &  0.5967  &  0.786  \\
72.0547   & 0.9054  &  0.791  & 72.1618   & 0.2438 &  0.646  &  72.2759  &  0.6043  &  0.772  \\
72.0566   & 0.9115  &  0.801  & 72.1645   & 0.2523 &  0.638  &  72.2782  &  0.6115  &  0.765  \\
72.0587   & 0.9181  &  0.818  & 72.1666   & 0.2590 &  0.641  &  72.2805  &  0.6188  &  0.756  \\
72.0612   & 0.9260  &  0.831  & 72.1686   & 0.2653 &  0.639  &  72.2831  &  0.6270  &  0.742  \\
72.0634   & 0.9329  &  0.846  & 72.1710   & 0.2729 &  0.645  &  72.2854  &  0.6343  &  0.733  \\
72.0654   & 0.9393  &  0.854  & 72.1732   & 0.2798 &  0.652  &  72.2878  &  0.6419  &  0.733  \\
72.0676   & 0.9462  &  0.867  & 72.1752   & 0.2861 &  0.654  &  72.2902  &  0.6494  &  0.727  \\
72.0705   & 0.9554  &  0.888  & 72.1772   & 0.2925 &  0.652  &           &          &         \\
\hline
\end{tabular}
\end{footnotesize}
\end{table}

\clearpage
\begin{table}
\begin{footnotesize}
\caption{CCD photometric data of GSC 02393-00680 in R band obtained
on January 6, 2008.}
\begin{tabular}{lllllllll}\hline\hline
JD (Hel.)    & Phase & $\Delta{m}$ & JD (Hel.)    & Phase &
$\Delta{m}$ & JD (Hel.)    & Phase & $\Delta{m}$ \\
2454400+   &             &            &2454400+     &            &
&2454400+ &              &            \\\hline
71.9753 & 0.6547 & 0.782 & 72.0735 & 0.9649 & 0.981 & 72.1781 & 0.2954 & 0.717 \\
71.9772 & 0.6607 & 0.774 & 72.0760 & 0.9728 & 0.993 & 72.1802 & 0.3020 & 0.723 \\
71.9791 & 0.6667 & 0.770 & 72.0780 & 0.9791 & 0.992 & 72.1821 & 0.3080 & 0.729 \\
71.9808 & 0.6721 & 0.758 & 72.0807 & 0.9877 & 1.005 & 72.1840 & 0.3140 & 0.727 \\
71.9832 & 0.6797 & 0.765 & 72.0835 & 0.9965 & 0.995 & 72.1865 & 0.3219 & 0.738 \\
71.9851 & 0.6857 & 0.758 & 72.0856 & 0.0032 & 1.006 & 72.1886 & 0.3286 & 0.744 \\
71.9870 & 0.6917 & 0.752 & 72.0875 & 0.0092 & 1.001 & 72.1910 & 0.3361 & 0.748 \\
71.9888 & 0.6973 & 0.752 & 72.0896 & 0.0158 & 0.999 & 72.1954 & 0.3500 & 0.764 \\
71.9908 & 0.7037 & 0.751 & 72.0919 & 0.0231 & 0.982 & 72.1975 & 0.3567 & 0.776 \\
71.9927 & 0.7097 & 0.736 & 72.0941 & 0.0300 & 0.985 & 72.1997 & 0.3636 & 0.781 \\
71.9945 & 0.7154 & 0.733 & 72.0959 & 0.0357 & 0.972 & 72.2020 & 0.3709 & 0.788 \\
71.9964 & 0.7214 & 0.740 & 72.0979 & 0.0420 & 0.965 & 72.2039 & 0.3769 & 0.799 \\
71.9982 & 0.7270 & 0.734 & 72.1000 & 0.0487 & 0.950 & 72.2061 & 0.3838 & 0.806 \\
72.0018 & 0.7384 & 0.716 & 72.1020 & 0.0550 & 0.937 & 72.2080 & 0.3898 & 0.814 \\
72.0036 & 0.7441 & 0.729 & 72.1040 & 0.0613 & 0.928 & 72.2107 & 0.3984 & 0.822 \\
72.0057 & 0.7507 & 0.734 & 72.1061 & 0.0679 & 0.905 & 72.2137 & 0.4079 & 0.833 \\
72.0076 & 0.7567 & 0.723 & 72.1084 & 0.0752 & 0.895 & 72.2158 & 0.4145 & 0.850 \\
72.0094 & 0.7624 & 0.732 & 72.1109 & 0.0831 & 0.884 & 72.2185 & 0.4230 & 0.856 \\
72.0112 & 0.7681 & 0.738 & 72.1132 & 0.0904 & 0.865 & 72.2210 & 0.4309 & 0.867 \\
72.0131 & 0.7741 & 0.737 & 72.1152 & 0.0967 & 0.855 & 72.2246 & 0.4423 & 0.891 \\
72.0151 & 0.7804 & 0.735 & 72.1173 & 0.1033 & 0.844 & 72.2271 & 0.4502 & 0.905 \\
72.0171 & 0.7868 & 0.738 & 72.1203 & 0.1128 & 0.831 & 72.2299 & 0.4590 & 0.923 \\
72.0190 & 0.7928 & 0.742 & 72.1229 & 0.1210 & 0.818 & 72.2324 & 0.4669 & 0.935 \\
72.0212 & 0.7997 & 0.747 & 72.1239 & 0.1242 & 0.820 & 72.2351 & 0.4755 & 0.951 \\
72.0233 & 0.8063 & 0.749 & 72.1246 & 0.1264 & 0.816 & 72.2373 & 0.4824 & 0.947 \\
72.0254 & 0.8130 & 0.755 & 72.1283 & 0.1381 & 0.793 & 72.2403 & 0.4919 & 0.950 \\
72.0275 & 0.8196 & 0.760 & 72.1303 & 0.1444 & 0.785 & 72.2429 & 0.5001 & 0.941 \\
72.0308 & 0.8300 & 0.771 & 72.1321 & 0.1501 & 0.779 & 72.2456 & 0.5086 & 0.956 \\
72.0330 & 0.8370 & 0.781 & 72.1340 & 0.1561 & 0.776 & 72.2477 & 0.5153 & 0.959 \\
72.0349 & 0.8430 & 0.779 & 72.1359 & 0.1621 & 0.771 & 72.2499 & 0.5222 & 0.949 \\
72.0374 & 0.8509 & 0.787 & 72.1393 & 0.1728 & 0.758 & 72.2521 & 0.5292 & 0.950 \\
72.0395 & 0.8575 & 0.793 & 72.1415 & 0.1798 & 0.746 & 72.2548 & 0.5377 & 0.950 \\
72.0415 & 0.8638 & 0.803 & 72.1435 & 0.1861 & 0.741 & 72.2570 & 0.5446 & 0.935 \\
72.0436 & 0.8705 & 0.806 & 72.1456 & 0.1927 & 0.737 & 72.2593 & 0.5519 & 0.919 \\
72.0456 & 0.8768 & 0.816 & 72.1479 & 0.2000 & 0.730 & 72.2617 & 0.5595 & 0.901 \\
72.0477 & 0.8834 & 0.820 & 72.1500 & 0.2066 & 0.727 & 72.2646 & 0.5687 & 0.888 \\
72.0497 & 0.8897 & 0.833 & 72.1521 & 0.2132 & 0.723 & 72.2675 & 0.5778 & 0.878 \\
72.0517 & 0.8961 & 0.843 & 72.1560 & 0.2256 & 0.721 & 72.2703 & 0.5867 & 0.867 \\
72.0536 & 0.9021 & 0.856 & 72.1580 & 0.2319 & 0.717 & 72.2724 & 0.5933 & 0.862 \\
72.0556 & 0.9084 & 0.863 & 72.1602 & 0.2388 & 0.716 & 72.2746 & 0.6002 & 0.842 \\
72.0577 & 0.9150 & 0.867 & 72.1629 & 0.2474 & 0.708 & 72.2771 & 0.6081 & 0.842 \\
72.0603 & 0.9232 & 0.892 & 72.1655 & 0.2556 & 0.693 & 72.2793 & 0.6151 & 0.833 \\
72.0622 & 0.9292 & 0.903 & 72.1676 & 0.2622 & 0.696 & 72.2819 & 0.6233 & 0.817 \\
72.0644 & 0.9362 & 0.919 & 72.1700 & 0.2698 & 0.706 & 72.2843 & 0.6309 & 0.813 \\
72.0663 & 0.9422 & 0.930 & 72.1722 & 0.2767 & 0.713 & 72.2867 & 0.6385 & 0.812 \\
72.0686 & 0.9495 & 0.938 & 72.1742 & 0.2831 & 0.705 & 72.2890 & 0.6457 & 0.794 \\
72.0715 & 0.9586 & 0.960 & 72.1763 & 0.2897 & 0.720 & 72.2913 & 0.6530 & 0.790 \\
\hline
\end{tabular}
\end{footnotesize}
\end{table}

\clearpage
\begin{table}
\caption{Times of light minimum of GSC 02393-00680.}
\begin{tabular}{lllrrrrl}
\hline\hline
JD.Hel.        &  Min.&  Method&  Error       &   E      &  $(O-C)_{1}$ & $(O-C)_{2}$ & Ref.$^{*}$\\
\hline
2453354.3960   &  I   &  CCD  &   $\pm0.0015$ & -1049    & 0.0005 &   0.0006   &  (1) \\
2453370.3815   &  II  &  CCD  &   $\pm0.0025$ & -998.5   & 0.0009 &   0.0010   &  (1) \\
2453373.3898   &  I   &  CCD  &   $\pm0.0022$ & -989     & 0.0021 &   0.0022   &  (1) \\
2453374.3372   &  I   &  CCD  &   $\pm0.0023$ & -986     & -0.0001&   0.0000   &  (1) \\
2453374.4992   &  II  &  CCD  &   $\pm0.0018$ & -985.5   & 0.0036 &   0.0037   &  (1) \\
2453375.2891   &  I   &  CCD  &   $\pm0.0020$ & -983     & 0.0022 &   0.0022   &  (1) \\
2453375.4405   &  II  &  CCD  &   $\pm0.0057$ & -982.5   & -0.0047&   -0.0046  &  (1) \\
2453376.2414   &  I   &   CCD &   $\pm0.0018$ & -980     & 0.0049 &   0.0049   &  (1) \\
2453377.3507   &  II  &   CCD &   $\pm0.0022$ & -976.5   & 0.0063 &   0.0064   &  (1)\\
2453377.5055   &  I   &   CCD &   $\pm0.0022$ & -976     & 0.0028 &   0.0029   &  (1) \\
2453683.4375   &  II  &   CCD &   $\pm0.0015$ & -9.5     & 0.0026 &   0.0018   &  (2) \\
2453683.5913   &  I   &   CCD &   $\pm0.0011$ & -9       & -0.0019&   -0.0027  &  (2) \\
2453686.4416   &  I   &   CCD &   $\pm0.0017$ & 0        & -0.0004&   -0.0012  &  (2) \\
2453686.5983   &  II  &   CCD &   $\pm0.0008$ & 0.5      & -0.0020&   -0.0028  &  (2) \\
2453694.3571   &  I   &   CCD &   $\pm0.0007$ & 25       & 0.0017 &   0.0009   &  (2) \\
2453694.5141   &  II  &   CCD &   $\pm0.0020$ & 25.5     & 0.0004 &   -0.0004  &  (2) \\
2453694.6702   &  I   &   CCD &   $\pm0.0003$ & 26       & -0.0017&   -0.0026  &  (2) \\
2453705.4330   &  I   &   CCD &   $\pm0.0010$ & 60       & -0.0012&   -0.0020  &  (2) \\
2453705.5913   &  II  &   CCD &   $\pm0.0010$ & 60.5     & -0.0011&   -0.0020  &  (2) \\
2453741.3591   &  II  &   CCD &   $\pm0.0012$ & 173.5    & -0.0019&   -0.0029  &  (2) \\
2453741.5188   &  I   &   CCD &   $\pm0.0009$ & 174      & -0.0005&   -0.0015  &  (2) \\
2453760.3483   &  II  &   CCD &   $\pm0.0010$ & 233.5    & -0.0049&   -0.0059  &  (2) \\
2453760.5127   &  I   &   CCD &   $\pm0.0009$ & 234      & 0.0012 &   0.0002   &  (2) \\
2453768.2638   &  II  &   CCD &   $\pm0.0007$ & 258.5    & -0.0028&   -0.0038  &  (2) \\
2453768.4251   &  I   &   CCD &   $\pm0.0016$ & 259      & 0.0002 &   -0.0008  &  (2) \\
2454130.3887   &  II  &   CCD &   $\pm0.0015$ & 1402.5   & 0.0047 &   0.0026   &  (2)\\
2454462.1148   &  II  &   CCD &   $\pm0.0010$ & 2450.5   & 0.0009 &   -0.0021  &  (3)\\
2454462.2716   &  I   &   CCD &   $\pm0.0010$ & 2451     & -0.0006&   -0.0036  &  (3)\\
2454472.0846   &  I   &   CCD &   $\pm0.0003$ & 2482     & -0.0002&   -0.0032  &  (3)\\
2454472.2446   &  II  &   CCD &   $\pm0.0004$ & 2482.5   & 0.0015 &   -0.0015  &  (3)\\
2454504.3716   &  I   &   CCD &   $\pm0.0010$ & 2584     & 0.0001 &   -0.0030  &  (4)\\
\hline
\end{tabular}
\noindent {\small $^{*}$ (1) Martignoni 2006; (2) Bl\"attler \&
Diethelm 2006; (3) this paper; (4) Diethelm 2008.\\}
\end{table}

\clearpage
\begin{table}
\caption{Photometric solutions for GSC 02393-00680.}
\begin{tabular}{lcl}
\hline
Parameters              &  Photometric elements  &  errors\\

                        &  spotted solutions   &        \\
\hline
$g_1=g_2$               &     0.32               & assumed\\
$A_1=A_2$               &     0.50               & assumed\\
$x_{1bolo},y_{1bolo}$   &0.185, 0.534            & assumed\\
$x_{2bolo},y_{2bolo}$   &0.245, 0.465            & assumed\\
$x_{1R},y_{1R}$         &0.105, 0.644            & assumed\\
$x_{2R},y_{2R}$         &0.218, 0.547            & assumed\\
$x_{1I},y_{1I}$         &0.029, 0.623            & assumed\\
$x_{1I},y_{1I}$         &0.115, 0.555            & assumed\\
$T_1$                   &     5860K              & assumed\\
$q$                     &     1.600            & $\pm0.004$\\
$\Omega_{in}$           &     4.6745             & --     \\
$\Omega_{out}$          &     4.0923             & --     \\
$T_2$                   &     5360K              & $\pm18$K \\
$i$                     &     60.071             & $\pm0.148$ \\
$L_1/(L_1+L_2)(R)$      &     0.4810             & $\pm0.0028$ \\
$L_1/(L_1+L_2)(I)$      &     0.4655             & $\pm0.0028$ \\
$\Omega_1=\Omega_2$     &     4.6453             & $\pm0.0034$\\
$r_1(pole)$             &     0.3204             & $\pm0.0008$\\
$r_1(side)$             &     0.3353             & $\pm0.0010$\\
$r_1(back)$             &     0.3692             & $\pm0.0016$\\
$r_2(pole)$             &     0.4219             & $\pm0.0008$\\
$r_2(side)$             &     0.4254             & $\pm0.0010$\\
$r_2(back)$             &     0.4521             & $\pm0.0013$\\
$f$                     &     $5.0\,\%$         & $\pm1.3\,\%$\\
$\theta$ ($^\circ$)     &     72.45                            \\
$\psi$($^\circ$)        &     283.28                          \\
$\Omega$($^\circ$)      &     60.46                           \\
$T_s/T_*$               &     0.8533                           \\
$\sum{(O-C)^2}$ &   0.0468400                         \\
\hline
\end{tabular}
\end{table}

\begin{table*}
\begin{minipage}{12cm}
\caption{The masses and orbital radii of the assumed third body in
GSC 02393-00680.}
\begin{tabular}{lll}
      \hline\hline
Parameters &  & Units\\
       \hline
$A_3$      & $0.0030(\pm0.0013)$   & d\\
$T_3$      & $1.92({\rm assumed})$ & yr\\
$e^{\prime}$ & $0({\rm assumed})$& ---\\
$a_{12}^{\prime}{\rm sin}i^{\prime}$ & $0.52(\pm0.23)$ & {\rm AU}\\
$f(m)$   & $0.0381$ & $M_{\odot}$\\
$m_{3}(i^{\prime}=90^{\circ})$  & $0.562(\pm0.007)$ & $M_{\odot}$ \\
$m_{3}(i^{\prime}=70^{\circ})$  & $0.607(\pm0.008)$ &  $M_{\odot}$ \\
$m_{3}(i^{\prime}=50^{\circ})$  & $0.784(\pm0.010)$ &  $M_{\odot}$ \\
$m_{3}(i^{\prime}=30^{\circ})$  & $1.400(\pm0.020)$ &  $M_{\odot}$ \\
$m_{3}(i^{\prime}=10^{\circ})$  & $9.836(\pm0.231)$ &  $M_{\odot}$ \\
$a_{3}(i^{\prime}=90^{\circ})$  & $2.00(\pm0.02)$ &  AU \\
$a_{3}(i^{\prime}=70^{\circ})$  & $2.01(\pm0.02)$ &  AU \\
$a_{3}(i^{\prime}=50^{\circ})$  & $2.06(\pm0.02)$ &  AU \\
$a_{3}(i^{\prime}=30^{\circ})$  & $2.23(\pm0.02)$ &  AU \\
$a_{3}(i^{\prime}=10^{\circ})$  & $3.48(\pm0.02) $ &  AU \\
        \hline\hline
\end{tabular}
\end{minipage}
\end{table*}

\begin{table*}
\caption{Some parameters of GSC 02393-00680 and XY Leo.}
\begin{tabular}{lccc}
      \hline\hline
Parameters &GSC 02393-00680 & XY Leo& Units\\
       \hline
$q_{ph}$ & 1.600   &1.64   & $M_2/M_1$\\
$T_h$    & 5860    &4850   & K\\
$T_c$    & 5360    &4524   & K\\
$i$      & 60.071  &68.2   & $\circ$\\
$f$      & $5.0\%$ &$2.44\%$&\\
$T_3$    & $1.92$  &$5.5$  &yr\\
$m_{3}$  & $0.67$  &$0.64$ & $M_{\odot}$ \\
$a_{3}$  & $2.03 $ &$3.8 $ & AU \\
        \hline
\end{tabular}
\noindent {\small
The values of $q_{ph}$, $T_h$, $T_c$, $i$, $f$ are
given by Djura\v{s}evi\'{c} et al. (2006); the $T_3$, $m_{3}$,
$a_{3}$ are given by Rucinski et al. (2007). \\}
\end{table*}

\begin{figure}
\begin{center}
\includegraphics[angle=0,scale=1 ]{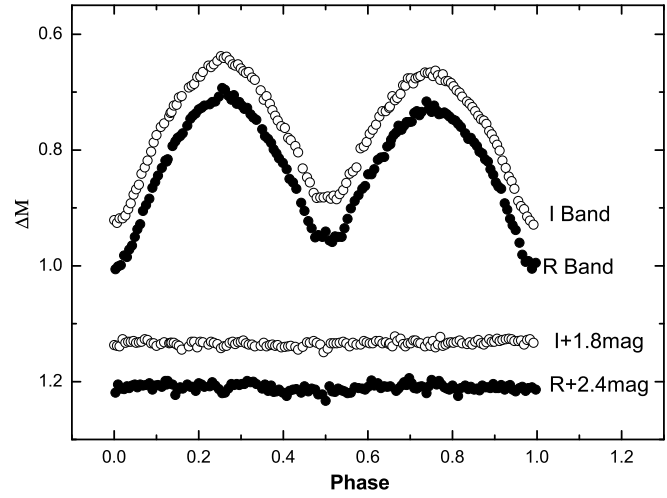}\caption{CCD photometric
light curves in $R$ and $I$ bands of GSC 02393-00680 obtained by 1m
reflecting telescope on January 6, 2008.} \end{center}
\end{figure}

\begin{figure}
\begin{center}
\includegraphics[angle=0,scale=1 ]{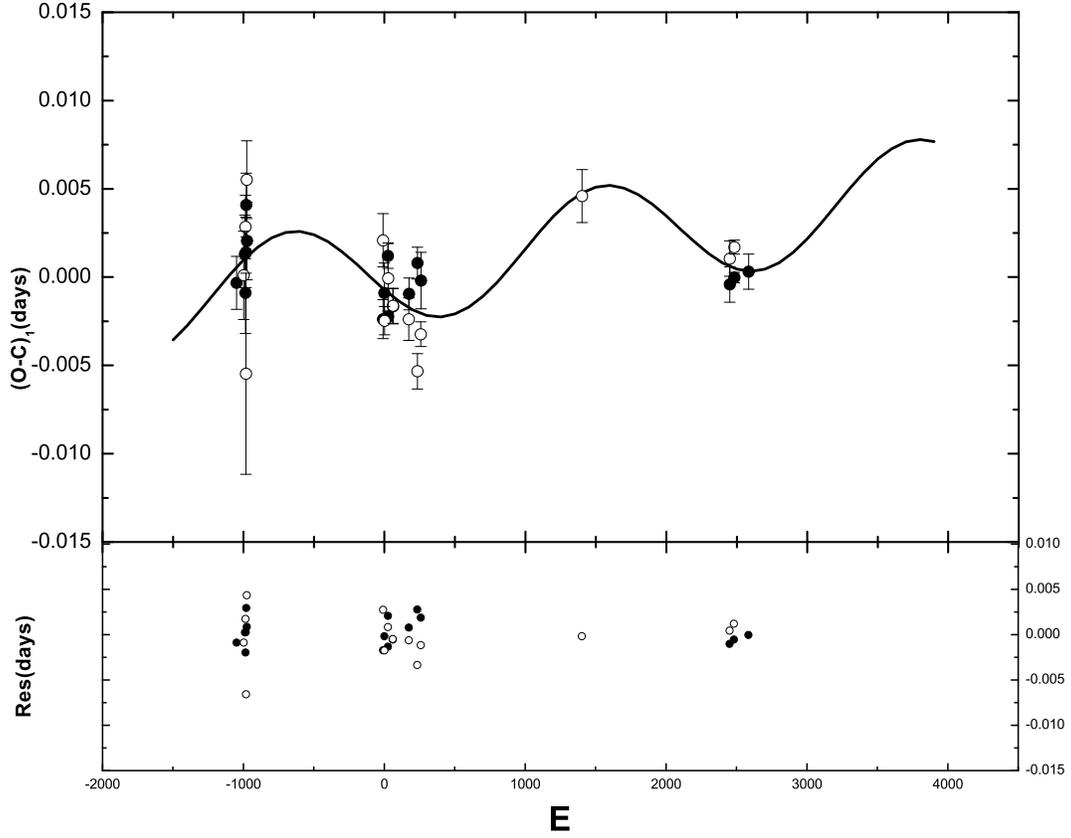}\caption{$(O-C)$ diagram
of GSC 02393-00680 formed by all available measurements. The
$(O-C)_1$ values were computed by using a newly determined linear
ephemeris (equation (1)). Solid cycles refer to the primary minimum
and open ones to the secondary minimum; solid line represents a
combination of a linear ephemeris and a cyclic variation. The
residuals yielded from equation (2) were displayed in the lower
panel of this figure.}\end{center}
\end{figure}

\begin{figure}
\begin{center}
\includegraphics[angle=0,scale=1 ]{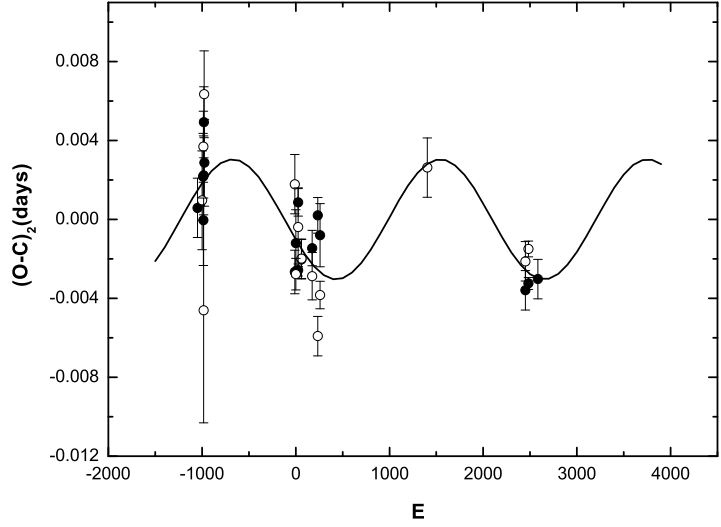}\caption{$(O-C)_{2}$
values for GSC 02393-00680 with respect to the cyclic ephemeris in
equation (2). Solid line refers to the theoretical orbit of an
assumed third body.} \end{center}
\end{figure}

\begin{figure}
\begin{center}
\includegraphics[angle=0,scale=1 ]{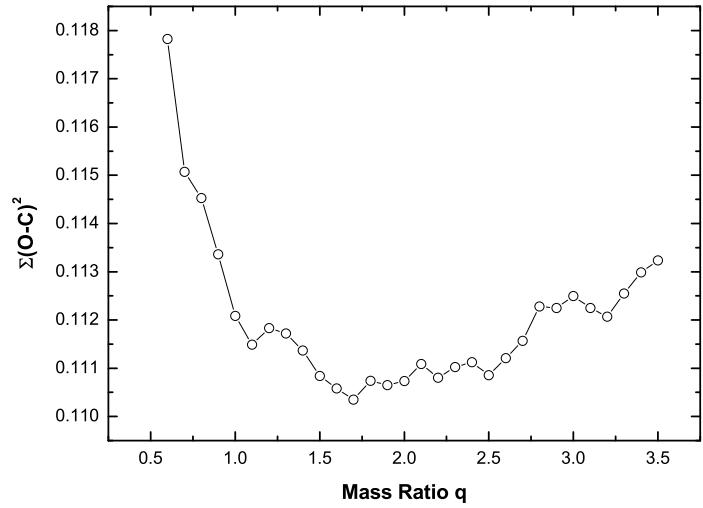}\caption{The relation
between q and $\Sigma$ for GSC 02393-00680.}\end{center}
\end{figure}

\begin{figure}
\begin{center}
\includegraphics[angle=0,scale=1 ]{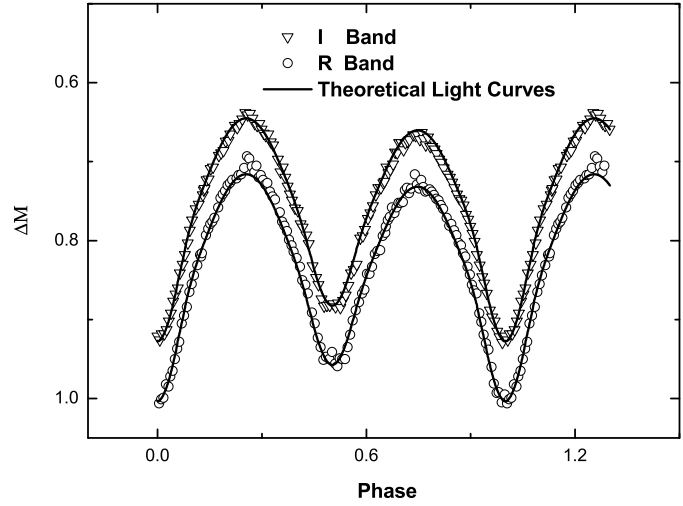} \caption{Observed
(circles) and theoretical (solid lines) light curves in $R$ and $I$
bands for GSC 02393-00680, with a dark spot in the more massive
component. } \end{center}
\end{figure}

\begin{figure}
\begin{center}
\includegraphics[angle=0,scale=1 ]{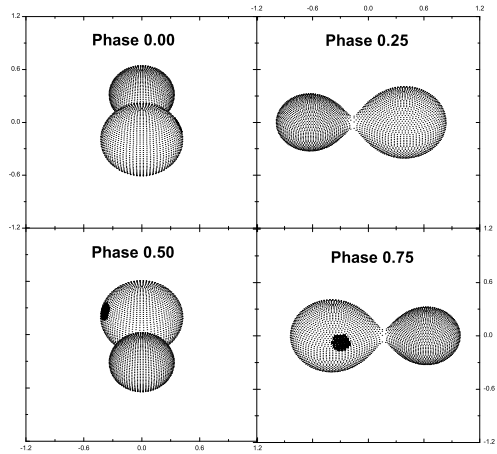} \caption{Geometrical
structure of the shallow contact binary GSC 02393-00680 with a dark
spot on the more massive component at phase 0.00, 0.25, 0.50 and
0.75.}\end{center}
\end{figure}

\end{document}